\documentclass[copyright,creativecommons]{eptcs}
\usepackage{multirow}
\usepackage{breakurl}   
\usepackage{verbatim}
\usepackage{color}
\usepackage{epsfig}
\usepackage{amsthm}
               
\newtheorem{mydef}{Definition}

\title{Modelling and Analysis of Biochemical Signalling\\ Pathway Cross-talk}
\author{Robin Donaldson, Muffy Calder\\
\institute{Department of Computing Science,\\ University of Glasgow\\
    Glasgow, UK }\\
    \email{radonald@dcs.gla.ac.uk}, \email{muffy@dcs.gla.ac.uk}\\
}	
      
\def\titlerunning{Modelling and Analysis of Biochemical Signalling Pathway Cross-talk}
\def\authorrunning{R. Donaldson \& M. Calder}
\begin{document}
\maketitle
            
\begin{abstract}   
Signalling pathways are abstractions that help life scientists structure the coordination of cellular
activity. Cross-talk between pathways accounts for many of the complex behaviours exhibited
by signalling pathways and is often critical in producing the correct signal-response relationship.
Formal models of signalling pathways and cross-talk in particular can aid understanding and drive
experimentation.  We define an approach to modelling based on the concept that a pathway is the (synchronising)
parallel composition of instances of generic modules (with internal and external labels).  Pathways are
then composed by (synchronising) parallel composition and renaming; different types of cross-talk
result from different combinations of synchronisation and renaming.
We define a number of generic modules in PRISM and five types of cross-talk:  signal flow, substrate availability, receptor
function, gene expression and intracellular communication. We show that Continuous Stochastic
Logic properties can both detect and distinguish the types of cross-talk.
The approach is illustrated with small examples and an analysis of the cross-talk between
the TGF-$\beta$/BMP, WNT and MAPK pathways.
\end{abstract}
                    
\section{Introduction}

Signalling pathways\footnote{we refer simply to pathways henceforth} are well-known abstractions that help life scientists structure the coordination of cellular 
activity.  Interaction between pathways, known as cross-talk, appears to have arisen for several reasons, for example to integrate signals, to produce a variety of responses to a signal, to reuse proteins between pathways; it 
 accounts for many of the complex signalling behaviours.  

This paper investigates modelling and analysis of pathway cross-talk; a key outcome is the suitability of process algebraic operators  to modelling cross-talk.  
  We define an approach to modelling based on the concept that a pathway is the (synchronising) parallel composition of instances of generic modules (with internal and external labels).  Pathways are then composed by (synchronising) parallel composition and renaming; different types of cross-talk result from different combinations of synchronisation and renaming.  
We use the PRISM modelling language and model checker. 
The contribution of the paper is the following:
\begin{itemize}

\item novel categorisation of types of cross-talk
\item modelling based on pathway module instantiation, internal/external reactions \&  synchronisation over subsets of (possibly renamed) external reactions
\item examples of modelling each type of  cross-talk 
\item detection and characterisation of cross-talk using Continuous Stochastic Logic 
\item application of approach to a case study: cross-talk between TGF-$\beta$/BMP, WNT and MAPK pathways.   
\end{itemize}

This paper is organised as follows.  The following section outlines
the background to signalling pathway cross-talk and defines a novel categorisation based on examples in the literature. 
In section \ref{model} we describe the modelling approach.  In section \ref{examples}
we show how each type of cross-talk is modelled for
an example pathway and how we can detect and characterise the cross-talk using CSL.
Section \ref{case-study} describes the case study, cross-talk between TGF-$\beta$/BMP, WNT and MAPK pathways, and gives some results.
Section \ref{discussion} contains a discussion of our overall approach and 
Section \ref{related} reviews related work. 
Conclusions and directions for future work are in section \ref{conclusions}.

\begin{figure}[ht]
\centering   
\begin{tabular}{ c  c }          
\includegraphics[trim=0in  0in 0in 0in, clip,  
angle=-90,width=2.5in]{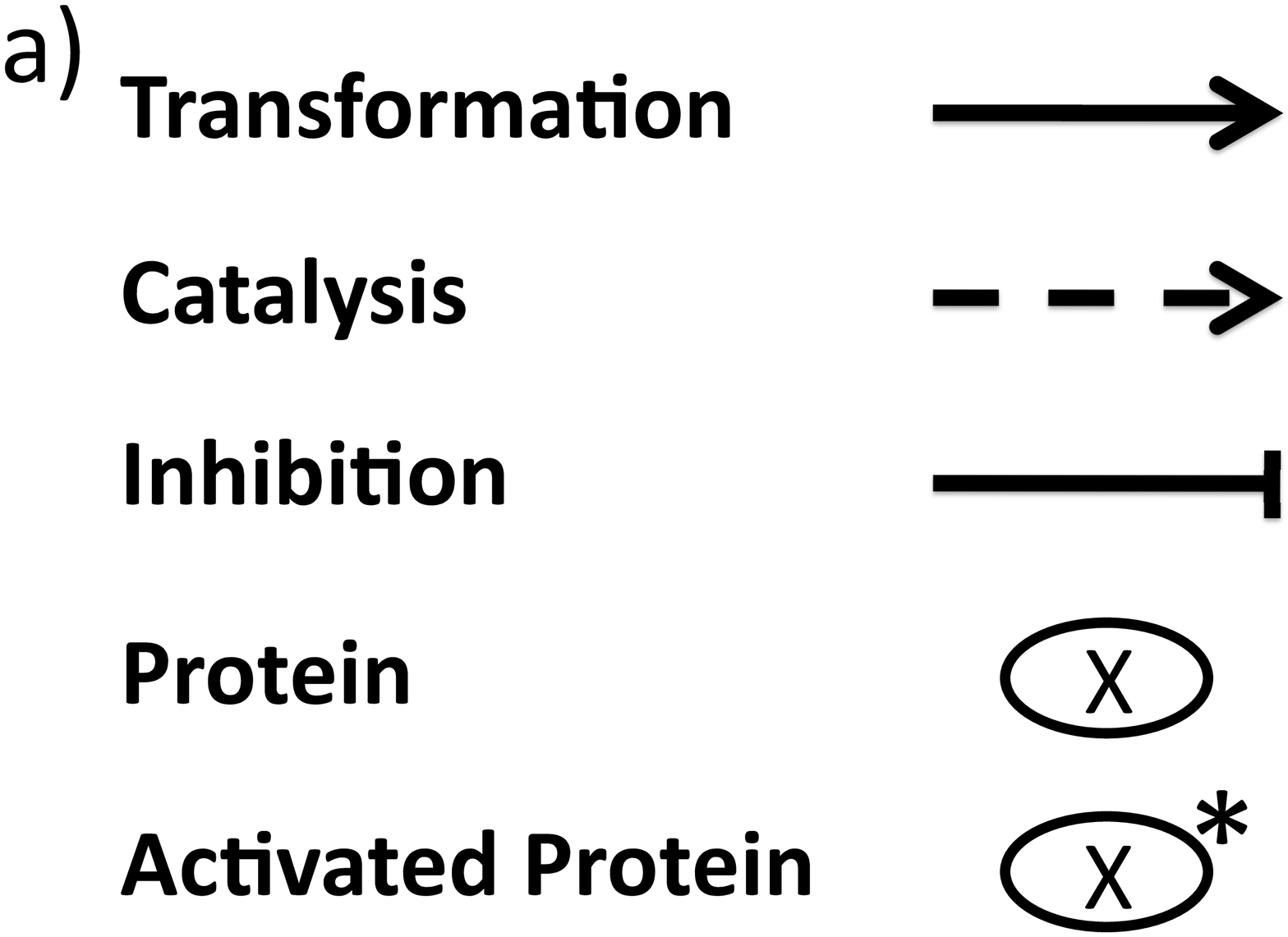} &                  
\includegraphics[trim=0in  0in 0in 0in, clip,  
angle=-90,width=2.5in]{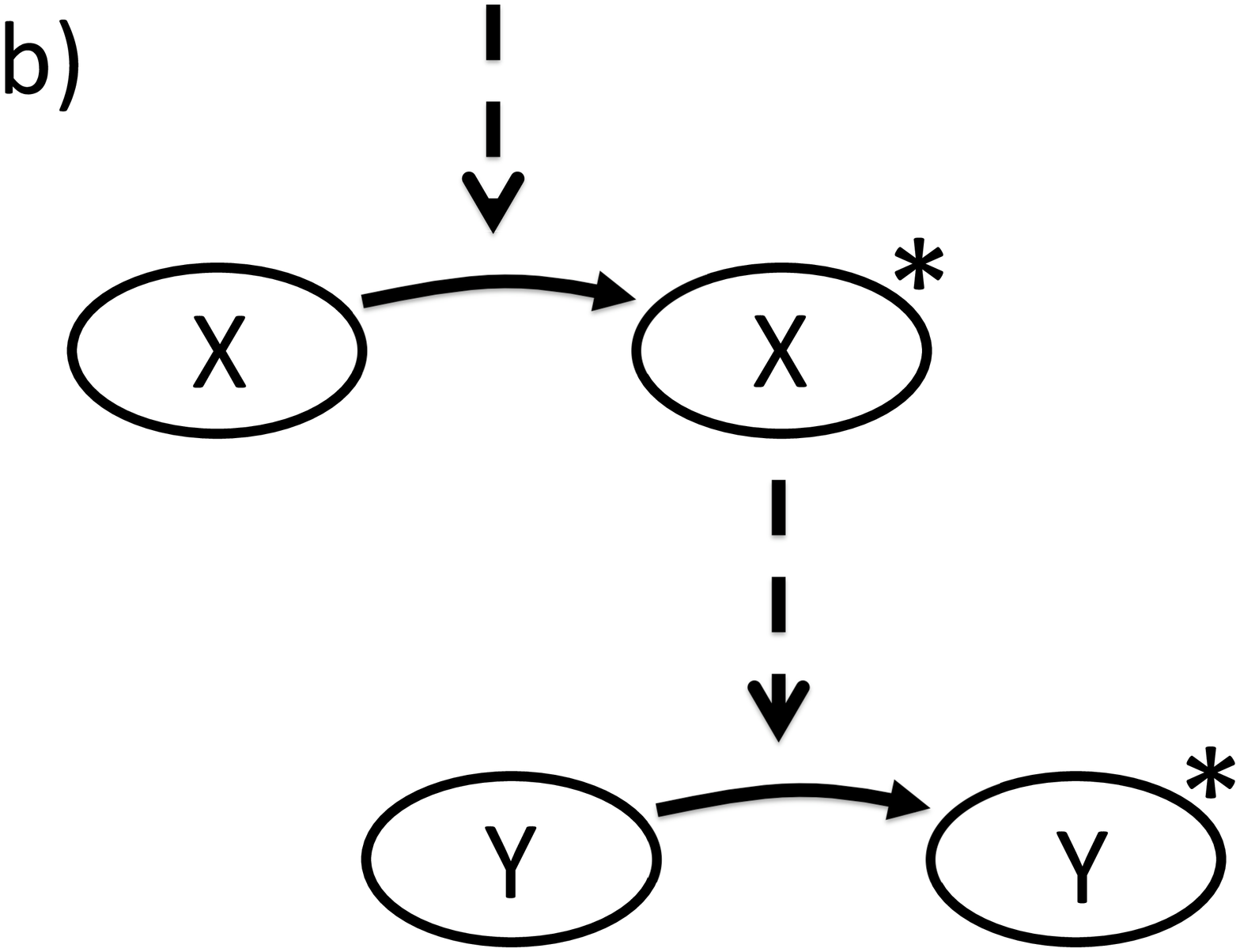} \\
\end{tabular}  
\caption{a) the notation used throughout this paper for 
arcs
and  nodes,  b) an example of a 2-stage signalling cascade in which the activated protein $X$  catalyses the activation of protein $Y$.
\label{cascadeFigure}} 
\end{figure}

Throughout the paper the following notation is used.  
Transformation (e.g. protein $X$ turns into protein $Y$) is denoted by a solid line with an arrow.  
Catalysis (increase in the rate of a reaction) is denoted by a dashed line with an arrow.  
Inhibition (decrease in the rate of a reaction) is denoted by a solid line with a blunt end.
Finally, we distinguish between inactive proteins and active proteins rather than the various mechanisms by which a protein changes state.  
An active protein is decorated with $*$, e.g  $X$ is  inactive  and $X*$ is active.
This notation is illustrated in Figure \ref{cascadeFigure}(a).

\section{Signalling Pathways \& Cross-talk}   
 \label{pathways}
Pathways are the mechanism by which a cell receives a signal and produces the appropriate cellular response.
A cell can respond to many different signals, and the same signalling pathway can produce different responses based on  different signals.  For example, the well-studied MAPK/ERK pathway can respond with cellular proliferation or cellular differentiation depending on the type of growth-factor present \cite{brightman2000}.
Many of the reactions involved in signalling pathways are enzyme  catalysed protein activations, often arranged in a ``signalling cascade''.  In such a cascade, the activated protein on one ``level'' is the enzyme for the activating reaction of the next ``level'', as shown in Figure \ref{cascadeFigure}(b).

Signalling pathways were first thought to be a linear series of reactions, but more recently, detailed understanding of these pathways shows that they are non-linear \cite{citeulike:479006}.  The series of reactions forming the  pathway can, for example,  diverge or interact upstream/downstream in the chain of reactions forming a  feedback/feedforward loop.
Laboratory techniques measuring the concentration of proteins or RNAs/DNAs in the cell under different conditions now enable biologists to build appropriate abstractions of pathways.  This is an incremental process further complicated by the lack of clear definition of what constitutes a single pathway. It is an inexact science to define the boundaries between pathways.  Often, the boundaries are simply drawn such that the abstraction of the  pathway can explain the biological data.

The term cross-talk was first applied to electronic circuits to describe a signal in one circuit having an undesired effect on another circuit \cite{catt1967}.  Cross-talk in this setting is a design flaw: the electronic circuit has been specified and built, and has resulted in an undesired interaction between signals, called ``signal interference''.   Biochemical cross-talk \cite{citeulike:1198248} is an interaction between signals flowing through two or more signalling pathways in a cell,  however, this 
is not necessarily indicative of signal interference.

\subsection{Types of Cross-talk}      
                        
Cross-talk can occur at all stages of signal propagation through a  pathway.                  
Although there is some discussion of  types of cross-talk \cite{guo2009}, there appears to be no universal 
categorisation in the literature. 
Here, we  define five  categories of cross-talk: signal flow, substrate availability, receptor function, gene expression and intracellular communication.
The five categories are illustrated in Figure \ref{crosstalkTypes} and are discussed in more detail below, with reference to an indicative example.
We note that four of the five categories are alluded to in  \cite{PA-07-058} but are not made specific.

\subsubsection{Signal Flow Cross-talk}  

Signal flow cross-talk between two  pathway occurs when a molecular species in one pathway affects the signal flow (rate of protein activation) in another pathway, shown in Figure  \ref{crosstalkTypes}(a).  The species  affects the signal flow by altering the rate(s) of the activation or deactivation reactions in the other pathway.  
Typically, this interaction  occurs in the cytoplasm and affects the rate of the downstream signal. 
For example, in \cite{citeulike:1198248} there is signal flow cross-talk between the MAPK and Integrin signalling pathways.  Activation of the Integrin pathway enhances signalling through the MAPK pathway by increased rate of activation of key proteins in the pathway.

\subsubsection{Substrate Availability Cross-talk} 
        
Substrate availability cross-talk occurs when two pathways compete for one or more common or homologous proteins (proteins that perform the same function),  shown in Figure  \ref{crosstalkTypes}(b).
For example 
\cite{citeulike:1097014} describes two pathways that compete for activation of the MAPK cascade.  The pathways share the MAPKKK protein STE11 and have homologous MAPKK and MAPK proteins.

\subsubsection{Receptor Function Cross-talk}    
        
Cross-talk can occur at the level of the pathway receptor, shown in Figure  \ref{crosstalkTypes}(c).  The receptor's ability to detect a ligand can be affected by cross-talk, for example, being  inhibited (thus slowing or blocking signal propagation) or  activated  (thus producing signal propagating in absence of the ligand).  
In \cite{estrogenReceptor} other signalling pathways can activate the estrogen receptor in the absence of the estrogen ligand.

\subsubsection{Gene Expression Cross-talk} 
         
Cross-talk can occur at the level of  gene expression within the nucleus, shown in Figure  \ref{crosstalkTypes}(d).  In this case, there is  an interaction between pathways in terms of  genes that will be expressed or repressed, for example, through activating/deactivating transcription factors (TFs) or increasing/reducing the number of TFs.    
In  
\cite{bosscher2006}  two 
 pathways  contain cross-talk within the nucleus.
One pathway contains a transcription factor called GR that resides outside the nucleus in its inactive state.  Upon activation (signalling), GR relocates to the nucleus and represses the transcription factor  NF-$\kappa$B that is activated through another pathway.

\subsubsection{Intracellular Communication Cross-talk}   
              
\label{cellCommSection}

\begin{figure}[ht!]
\centering

\begin{tabular}{cc}                            
\includegraphics[trim=.5in  .5in .5in .5in, clip,
	angle=-90,width=2.5in]{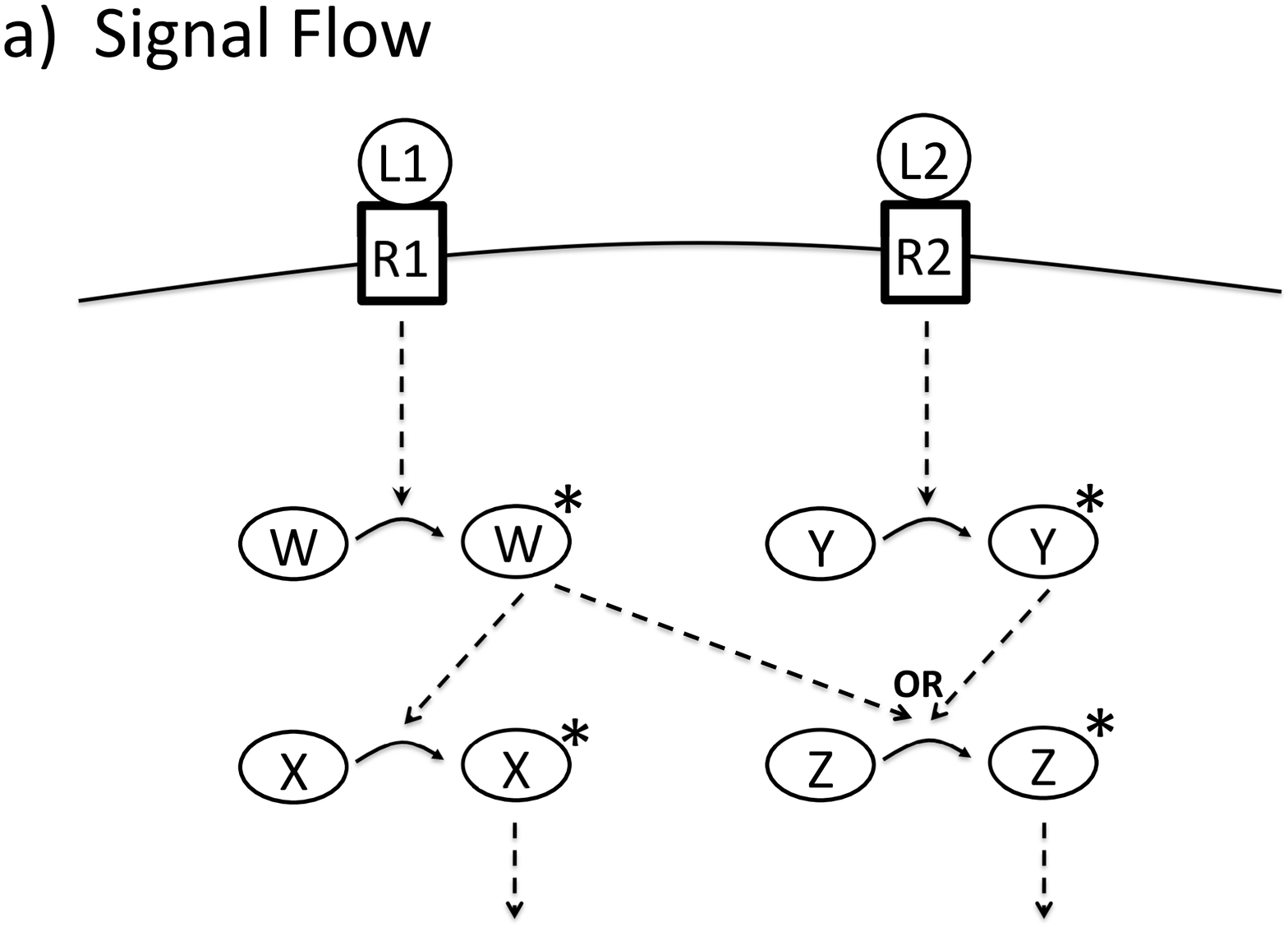}
&
	\includegraphics[trim=.5in  .5in .5in .5in, clip,  
	angle=-90,width=2.5in]{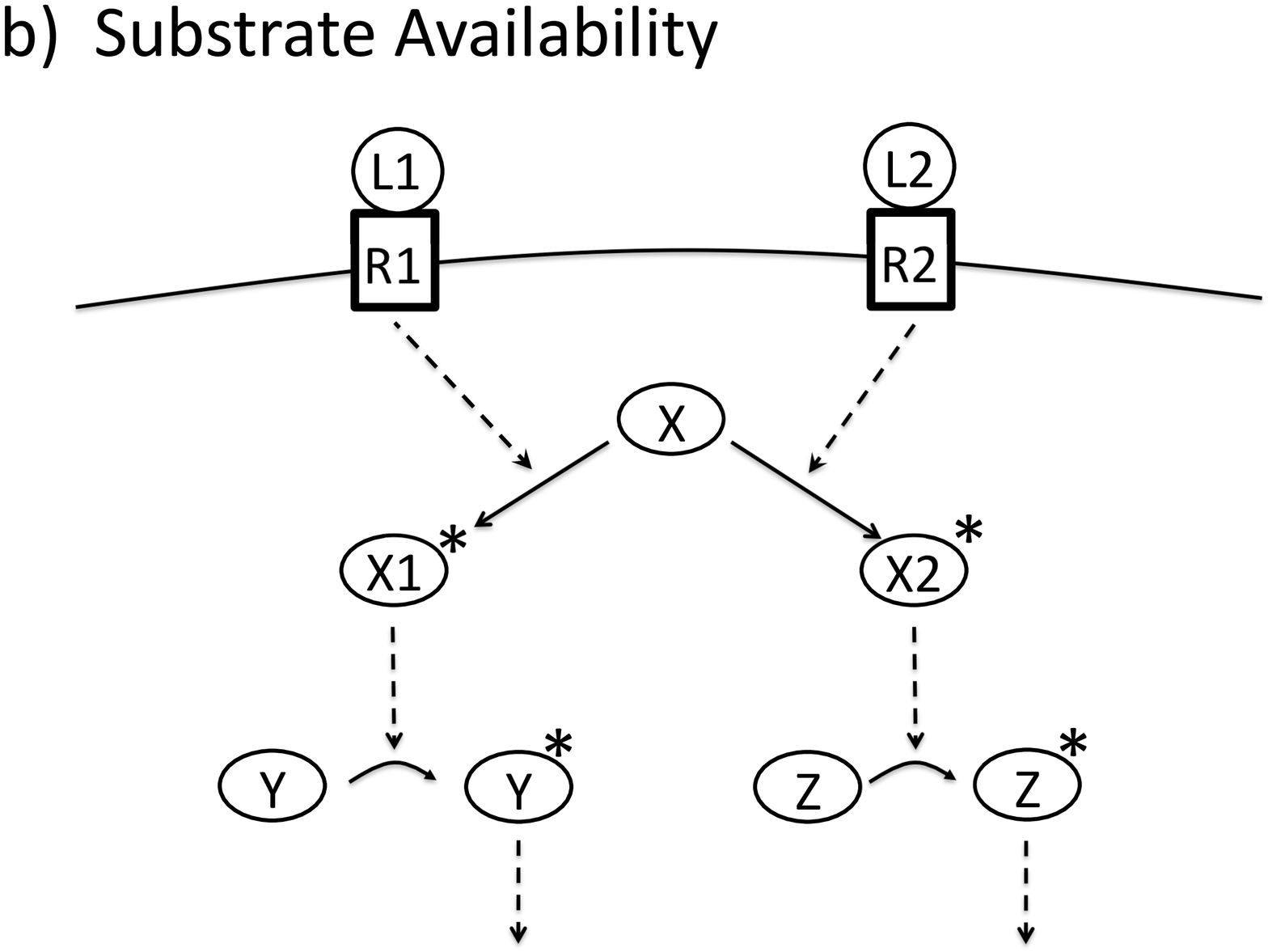}
\\
	\includegraphics[trim=.5in  .5in .5in .5in, clip,
	angle=-90,width=2.5in]{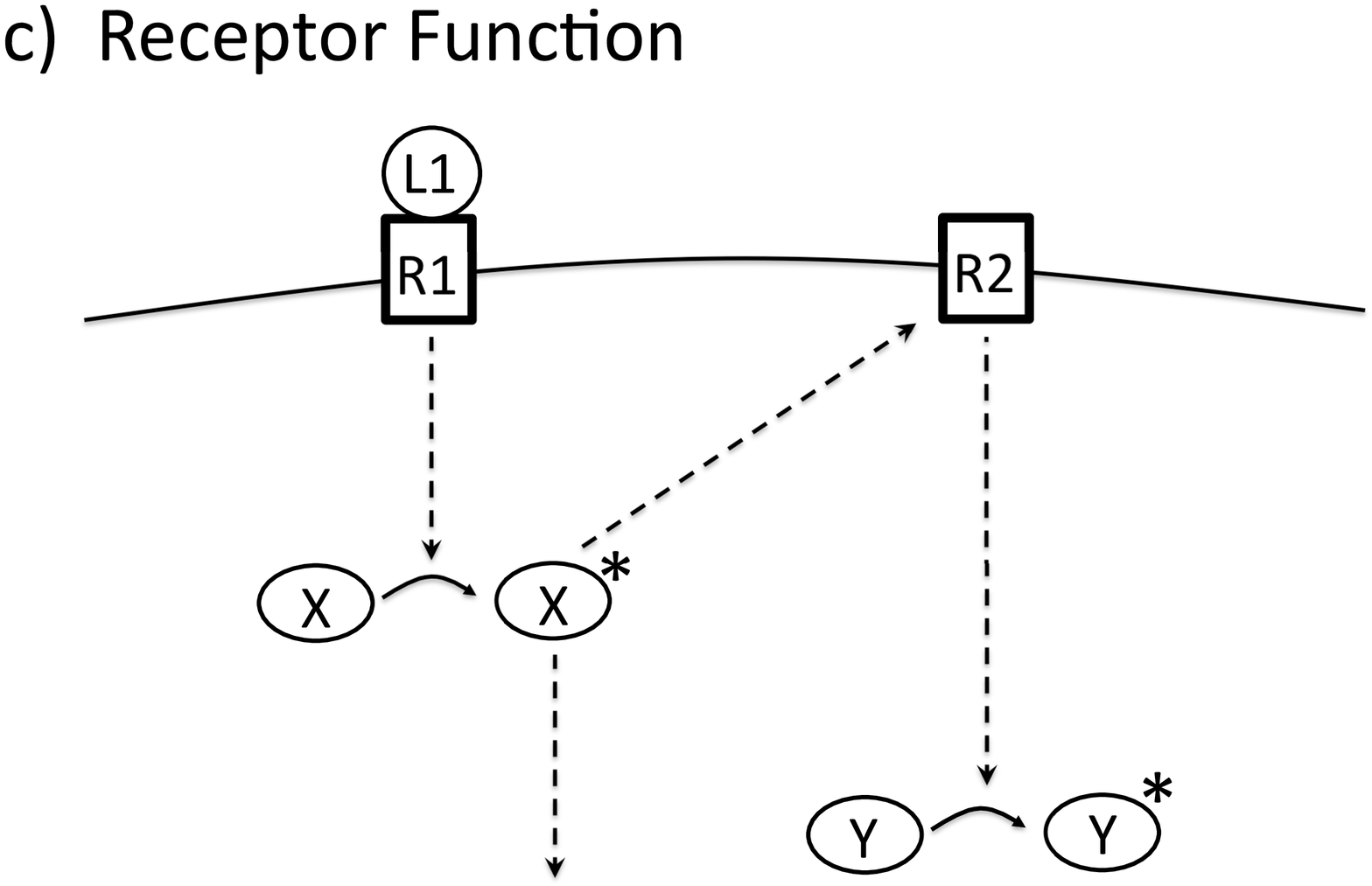}
&
	\includegraphics[trim=.5in  .5in .5in .5in, clip,
	angle=-90,width=2.5in]{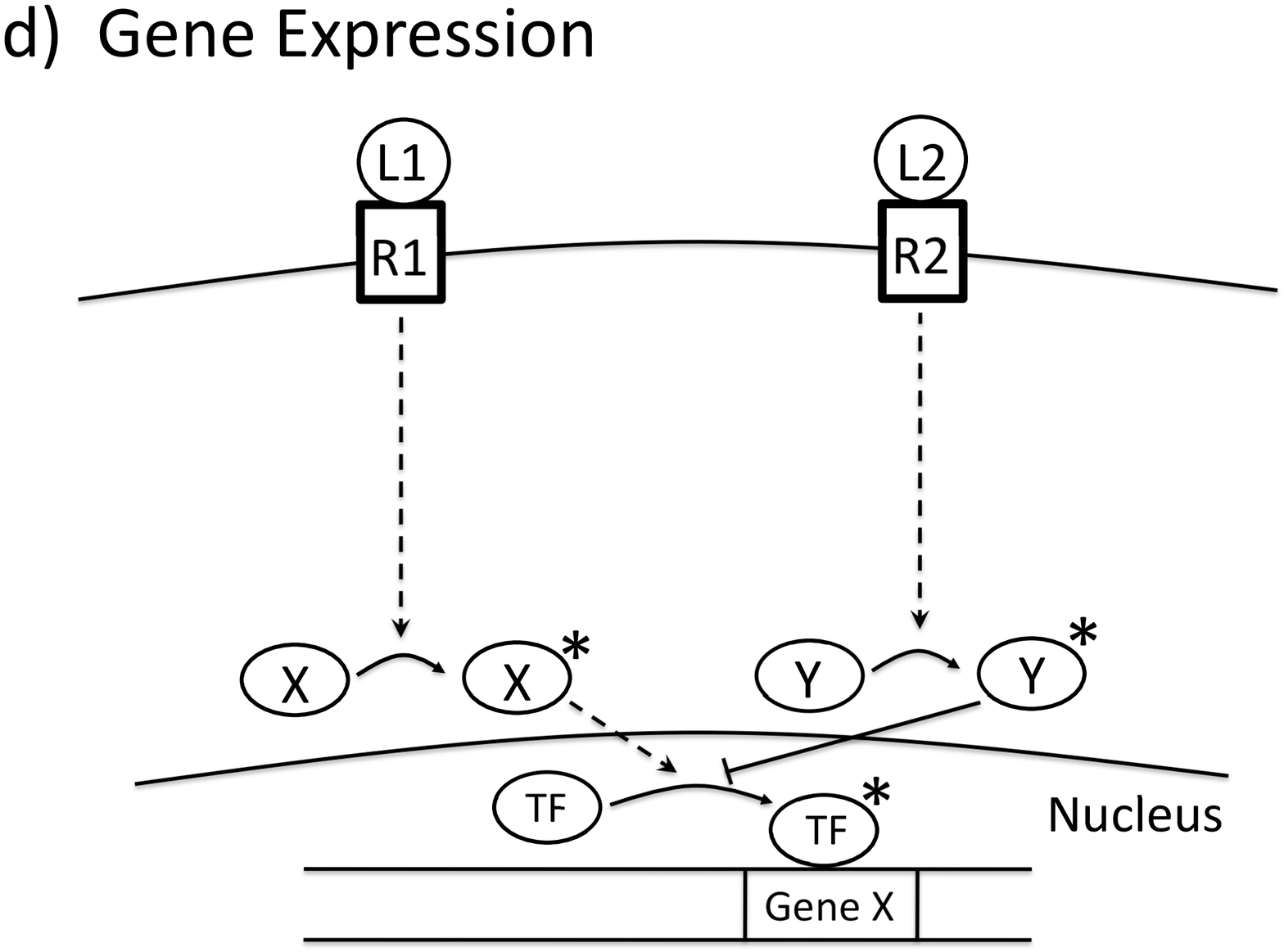}
\\
\end{tabular}
       
	\includegraphics[trim=.5in  .5in .5in .5in, clip,
angle=-90,width=2.5in]{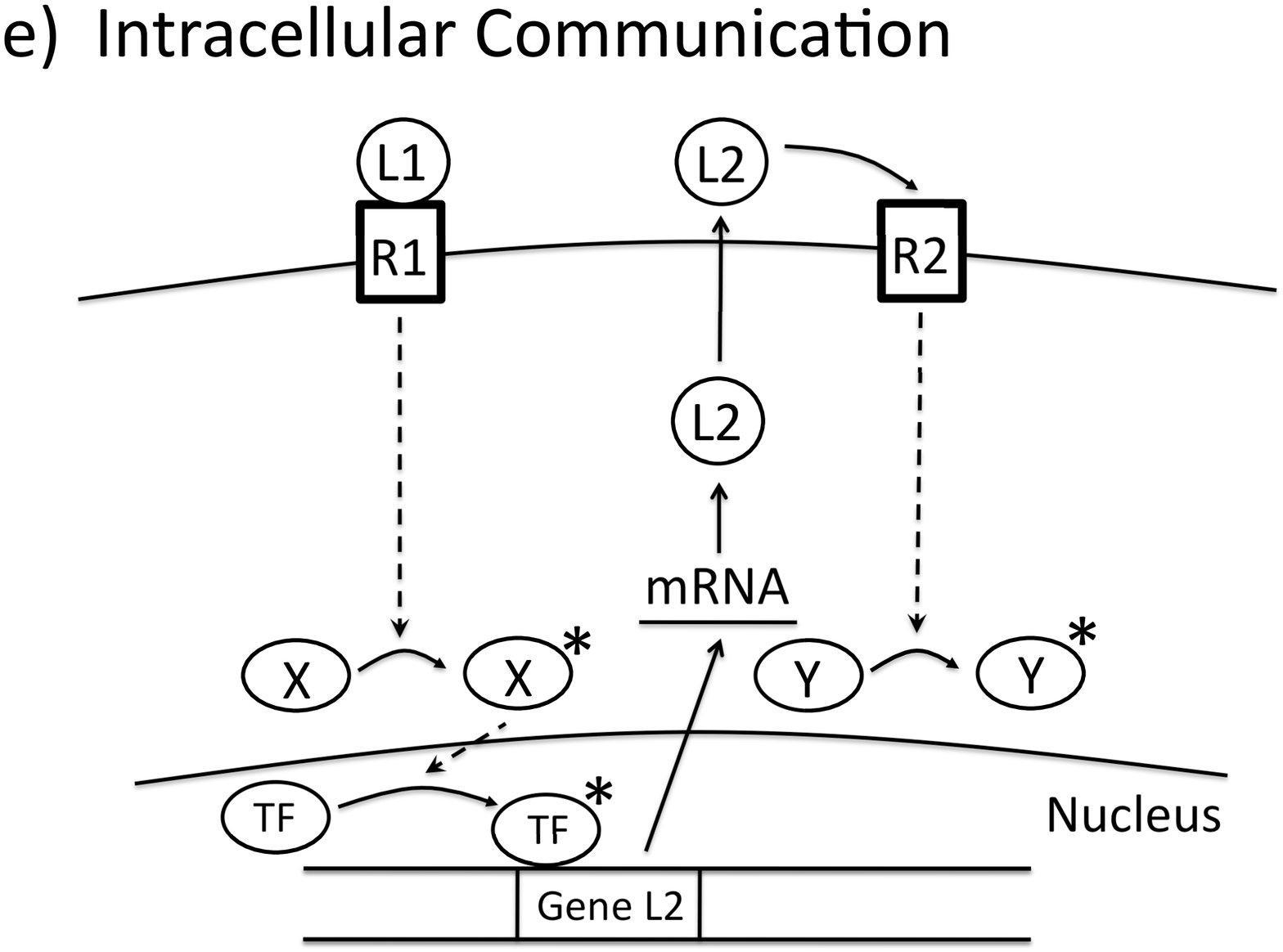}
        
\caption{An example of each of the five types of cross-talk:        
a) a pathway up-regulates  signal flow through another pathway, 
b) two pathways compete for a protein,  
c) a pathway activates the receptor of another pathway in the absence of a ligand,  
d) two pathways have conflicting  transcriptional responses,  
e) a pathway releases a ligand for another pathway.  The following notation is used:  L1/L2 - ligands, R1/R2 - receptors, W/X/Y/Z - Proteins, TF - Transcription Factor.
\label{crosstalkTypes}} 
\end{figure}

Finally, signalling pathways can cross-talk using the  less direct manner of intracellular communication, shown in Figure  \ref{crosstalkTypes}(e).
Instead of physical interaction between the proteins that comprise the pathways, one pathway can release a ligand that activates another pathway.           
In \cite{guo2009} the TGF-$\beta$/BMP and WNT pathways reciprocally regulate the production of their ligands.
There is some contention in the literature as to whether this is genuine cross-talk:  the interaction is less-direct than other types of cross-talk and involves lengthy processes such as gene expression and ligand excretion.

\section{Modelling}
\label{model}   
     
Our aim is to build models of pathways and their cross-talk in a modular fashion, whilst demonstrating our cross-talk categorisation.  

The literature contains many examples of modelling biological systems based on an unstructured set of equations, for example  Ordinary Differential Equations (ODEs) \cite{orton2005}.  
However,  flat equations have  disadvantages such as lack of structure and modularity -- there is no accepted way of dealing the with composition of these models.
Several  formal languages from Computer Science have well-understood notions of structure.  

We choose a state-based formalism, the PRISM modelling language \cite{www:prism}, with modules, renaming and synchronous communication between modules.  The semantics is given by continuous-time Markov chains (CTMCs), which can be analysed using the PRISM model checker with Continuous Stochastic Logic (CSL).  We give a brief overview of the language.  

The language is based on ``reactive modules'', 
each of which 
can contain local variables. The variables are updated by the execution of commands that have the following syntax, \linebreak
\verb=[label] guard -> rate:update_statement=.
Commands are only 
executable when their \verb=guard= becomes true, and any \verb=label= synchronises as required.  The \verb=rate= is used to build 
the underlying CTMC, providing both the probability and timing information of the state after \verb=update_statement= executes.

Modules can be composed concurrently, synchronising (multiway) on the commands whose labels occur in the synchronisation set, $L$. 
For example, given modules $M1$ and $M2$, and set of labels $L$, $M1~|[L]|~M2$ denotes the concurrent composition of $M1$ and $M2$, synchronising on all labels in L.  
If the label set is omitted, $M1~||~M2$, then $M1$ synchronises with $M2$ on the intersection of labels 
occurring in $M1$ and $M2$.
PRISM also allows renaming of labels, denoted thus $M1~\{old\_label~\leftarrow~new\_label\}$, and hiding, denoted thus $M~/~\{label1,~\ldots,~labeln\}$. Hidden labels are not available for synchronisation.

\subsection{Generic Modules}

\begin{figure}[ht]
\centering            
\includegraphics[trim=.5in 1in 3.4in 1in, clip,  
angle=-90,width=4.8in]{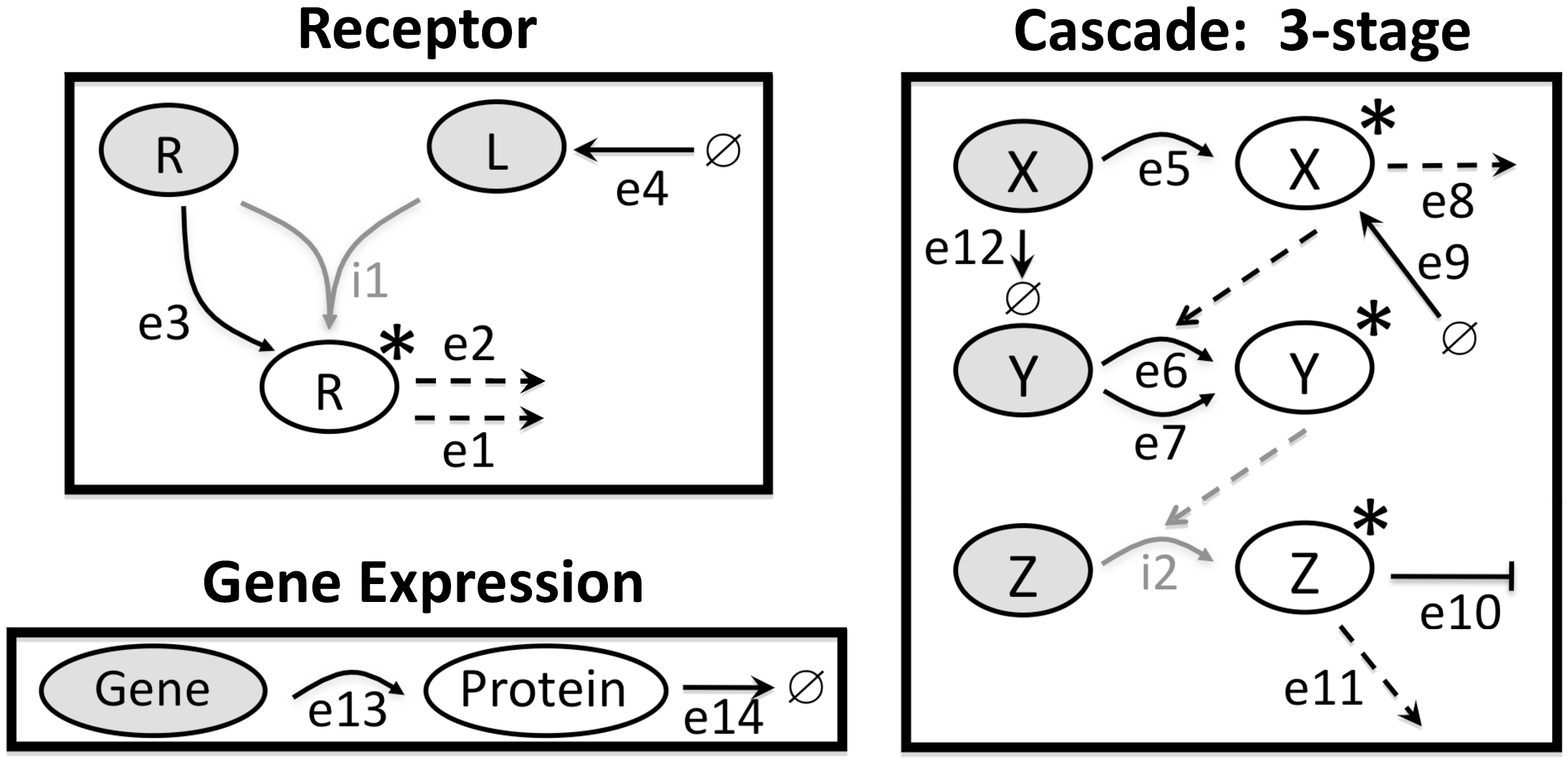}
\caption{Three generic pathway modules:  receptor, 3-stage cascade and gene expression.  
There are 14 external reactions ($e1,\ldots,e14$) denoted by black arcs, 2 internal reactions ($i1,i2$) denoted by gray arcs and 6 species that are  present in the initial state denoted by shaded ellipses.  
\label{modulesFigure}} 
\end{figure}

We define a  {\it pathway module} to be a behavioural pattern within a pathway. 
For example, commonly occurring pathway modules are: receptor, 3-stage cascade and gene expression (Figure \ref{modulesFigure}).

We represent these by generic modules in PRISM as follows.
We adopt the reagent-centric modelling style, as first presented in \cite{2005cbio}. 
Local variables contain the concentration of proteins in a particular state (e.g. active, inactive). 
Labelled commands change protein concentration according to biochemical reactions.
Reactions are considered to be external or internal.
The former denote behaviour that can coordinate (or be coordinated with)  behaviour in other modules, they are available for synchronisation. The latter reactions are hidden and are not available for synchronisation.   
Concentration is modelled by discrete, abstract {\it levels}, as defined in \cite{ctmcLevels}. 
Here, we consider strong abstractions, usually two levels \{0,1\}, and unit reaction rates. 
Due to the ligand production reaction $e4$, ligand has 
three levels \{0,1,2\}, and  the rate of ligand-receptor binding reflects the  level of the ligand (law of mass-action).

The PRISM modules for receptor,  3-stage cascade and gene expression are the following.

\begin{footnotesize}
\begin{verbatim}
module Receptor
   R : [0..1] init 1;    L1 : [0..2] init 1;    R1Active : [0..1] init 0;
 
   [i1_1] R1 = 1 & L1 >= 1 & R1Active = 0 -> L1:(R1' = 0) & (L1' = 0) & (R1Active' = 1);
   [e1_1] R1Active = 1 -> 1:(R1Active' = R1Active);
   [e2_1] R1Active = 1 -> 1:(R1Active' = R1Active);
   [e3_1] R1 = 1 & R1Active = 0 -> 1:(R1' = 0) & (R1Active' = 1);
   [e4_1] L1 < 2 -> 1:(L1' = L1 + 1);
endmodule

module Cascade3
   X1Inactive : [0..1] init 1;    X1Active : [0..1] init 0;    Y1Inactive : [0..1] init 1;
   Y1Active : [0..1] init 0;      Z1Inactive : [0..1] init 1;  Z1Active : [0..1] init 0;
 
   [e5_1] X1Inactive = 1 & X1Active = 0 -> 1:(X1Inactive' = 0) & (X1Active' = 1);
   [e6_1] Y1Inactive = 1 & Y1Active = 0 & X1Active = 1 -> 1:(Y1Inactive' = 0) & (Y1Active' = 1); 
   [e7_1] Y1Inactive = 1 & Y1Active = 0 -> 1:(Y1Inactive' = 0) & (Y1Active' = 1); 
   [i2_1] Z1Inactive = 1 & Z1Active = 0 & Y1Active = 1 -> 1:(Z1Inactive' = 0) & (Z1Active' = 1);
   [e8_1] X1Active = 1 -> 1:(X1Active' = X1Active);
   [e9_1] X1Active = 0 -> 1:(X1Active' = 1);
   [e10_1] Z1Active = 0 -> 1:(Z1Active' = Z1Active);
   [e11_1] Z1Active = 1 -> 1:(Z1Active' = Z1Active);
   [e12_1] X1Inactive = 1 -> 1:(X1Inactive' = 0);
endmodule

module GeneExpression
   Gene1 : [0..1] init 1;    Protein1 : [0..1] init 0;
  
   [e13_1] Gene1 = 1 & Protein1 = 0 -> 1:(Gene1' = 0) & (Protein1' = 1);
   [e14_1] Protein1 = 1 -> 1:(Protein1' = 0);
endmodule
\end{verbatim}
\end{footnotesize}

We will treat these modules as {\it generic}, that is, we instantiate them (strictly, duplicate and rename in PRISM) for multiple occurrences.  For brevity, we adopt the following convention. For generic module $M$, $M_{i}$ denotes an {\it instance} of $M$ with every variable and label renamed by an indexed form. For example, variable $v$ becomes $v_{1}$ in module $M_{1}$.

\subsection{Pathways}

A pathway is a parallel composition of instances of the generic modules, renaming labels to coordinate synchronisation within the pathway.

\begin{mydef}
Let G  be a set of generic modules.  A pathway $P$ has the form $f_{1}X_{1}~|[L_1]|~\ldots~|[L_{n-1}]|~f_{n}X_{n}$  where 
$X_{1}~\ldots~X_{n}$ are instances of modules in $G$, $f_{1}~\ldots ~f_{n}$ are compositions of renamings and hidings and $L_1~\ldots~L_{n-1}$ are labels.  
\end{mydef}

As an example, consider the expression of a pathway  comprising the active receptor catalysing the activation of protein X and active protein Z catalysing the expression of Gene.  This is defined by:  

$P_{1} ~ = ~ Receptor_{1} ~ / ~ \{i1_{1}\} ~ \{e1_{1} ~ \leftarrow ~e5_{1}\} ~
               |[e5_{1}]|  ~ Cascade3_{1}~/~\{i2_{1}\}~\{e11_{1}~\leftarrow ~e13_{1}\}~$\\
\indent\indent\indent
$               |[e13_{1}]|  ~ GeneExpression_{1} $

The $Receptor_{1}$ and $Cascade3_{1}$ modules synchronise on $e5_{1}$, and $Cascade3_{1}$ and $GeneExpression_{1}$ synchronise on $e13_{1}$; in both cases these are the only external labels that occur in both modules.  
Internal reactions  $i1_{1}$ and $i2_{1}$ are hidden using the $/$ operator.

\subsection{Composition of Pathways}

In a similar way, pathways can be composed, synchronising on  external labels.

\begin{mydef}
Given two  pathways $P_{1}$  and $P_{2}$, with   sets of external labels $ext(P_1)$ and $ext(P_2)$ resp., 
if $ext(P_{1})~\cap~ext(P_{2})  = \{\}$ then the pathways are independent, otherwise there is  crosstalk. 
\end{mydef}

\begin{figure}[ht]
\centering         
\includegraphics[trim=0in 0in 0in 0in, clip,  
angle=-90,width=5.5in]{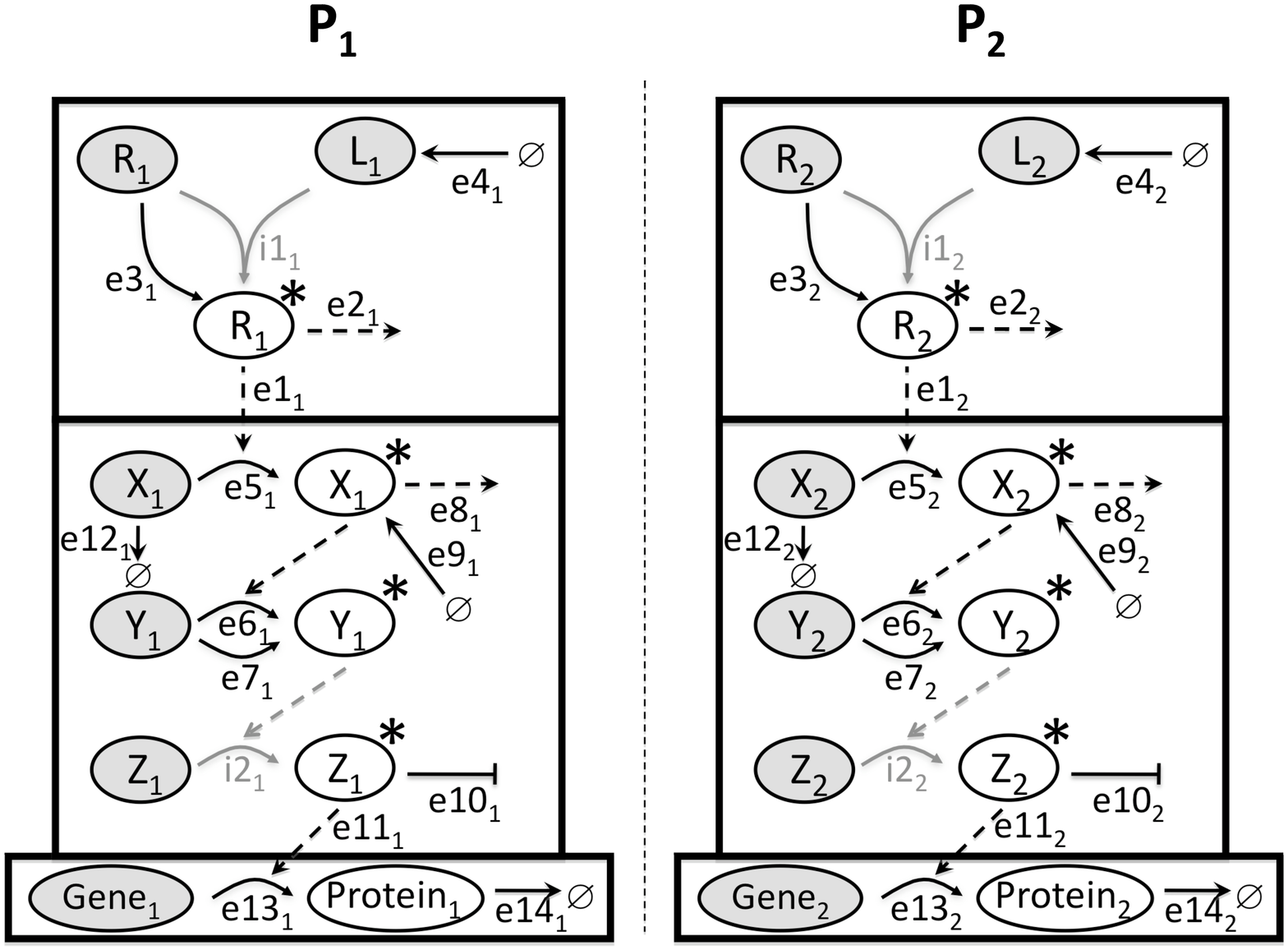}
\caption{The two pathways $P_1$ and $P_2$ each comprising three instances of the generic pathway modules receptor, 3-stage cascade and gene expression.  External reactions are denoted by black arcs, internal reactions by gray arcs
and species that are  present in the initial state  by shaded ellipses.
\label{PathwayDiagram}} 
\end{figure}

As an example of independent pathways, consider $P_{1}$ and $P_{2}$ (shown in Figure \ref{PathwayDiagram}), where

$P_{2} ~ = ~ Receptor_{2} ~ / ~ \{i1_{2}\} ~ \{e1_{2} ~ \leftarrow ~e5_{2}\} ~
               |[e5_{2}]|  ~ Cascade3_{2}~/~\{i2_{2}\}~\{e11_{2}~\leftarrow ~e13_{2}\}~$\\
\indent\indent\indent
$               |[e13_{2}]|  ~ GeneExpression_{2} $

We synchronise  $P_{1}$ and $P_{2}$ over the unused
 external reactions, i.e. the reactions that are otherwise never involved in synchronisation.  In this case,  
$P_{1}~|[U]|~P_{2}$ where $U=$ $\{e2_1,~e3_1,~e4_1,~e7_1,~e8_1,~e9_1,~e10_1,$      $~e12_1,~e14_1,~e2_2,~e3_2,~e4_2,~e7_2,~e8_2,~e9_2,~e10_2,~e12_2,~e14_2\}$.
Since the labels in $U$ occur in only one pathway they cannot synchronise and  their corresponding actions  never execute.

$P_{1}$ and $P_{2}$ are independent pathways, since $ext(P_{1})~\cap~ext(P_{2}) = \{\}$.     
We can also demonstrate independence through CSL properties. For example, the following CSL property expresses that it is not possible to activate $P_1$ without activating receptor $R_1$ or activate $P_2$ without activating  receptor $R_2$.

$P_{1}~|[U]|~P_{2}~\models~\mathbf{P_{\leq 0}}~[~F(RActive_1~=~0~\wedge~Protein_1~=~1)~\vee~
F(RActive_2~=~0~\wedge~Protein_2~=~1)~]$

In the next section we consider the case where the intersection is not empty; specifically, we consider examples of the five possible types of cross-talk between $P_{1}$ and $P_{2}$.

\section{Cross-talk models}  
\label{examples}

We now consider the five possible types of cross-talk between $P_{1}$ and $P_{2}$.  
In the previous section  we synchronised $P_{1}$ and $P_{2}$ over $U$, the set of unused reactions.
We now synchronise over $E$ and $U$,  $~P_{1}~|[E,~U]|~P_{2}$,
where $E=ext(P_{1})~\cap~ext(P_{2})$, i.e. the common external labels between $P_{1}$ and $P_{2}$, which may involve renaming. 
For each type we give a general description and then an explicit example.

We say that there is competition for a protein if the both pathways modify the protein such that they become unavailable to the other pathway.
An enzyme that is shared between two pathways  could be considered  competition, however the duration of the  enzyme-substrate complex is extremely short.  As such, we ignore this step in our approach.

\paragraph{Signal Flow Cross-talk}   

$P_{1}~|[E,~U]|~P_{2}$
has signal flow cross-talk if $E$ only contains labels from two cascade modules or labels  from a cascade module and catalysis/inhibition labels from a receptor module, in either case not setting up competition for a protein between the pathways.  Example: provide
 an additional route to the activation of $Y_{1}$ through the  $X_2*$ enzyme.
Synchronise $e8_2$, the enzymatic activity of $X_2*$, with $e7_1$ the alternative route to activate $Y_1$,  by renaming $e8_2$ to $e7_1$ and synchronising between pathways on $e7_1$.

$P_{1}~|[e7_1,~U]|~P_{2}~\{e8_2~\leftarrow~e7_1\}$

 where $U=\{e2_1,~e3_1,~e4_1,~e8_1,~e9_1,~e10_1,~e12_1,~e14_1,~e2_2,~e3_2,~e4_2,~e7_2,~e9_2,~e10_2,~e12_2,~e14_2\}$.

\paragraph{Substrate Availability Cross-talk}   

$P_{1}~|[E,~U]|~P_{2}$
has substrate availability cross-talk if $E$ only contains labels from  cascade modules or labels  from a cascade module and catalysis/inhibition labels from a receptor module, in either case setting up competition for a protein between the pathways.
Example: make both pathways compete for the activation of protein $X_1$ catalysed by  their respective receptors $R1*$ and $R2*$.
Within $P_2$, synchronise  $e2_2$, the enzymatic activity of $R_2*$, with $e9_2$, the production of $X_2*$.
Between pathways, synchronise $e12_1$, the degradation of $X_1$, with the $e9_2$.
This produces a new reaction, $X_1~\rightarrow~X_2*$ with $R_2*$ as the enzyme.
Also, synchronise between pathways on $e5_2$ to block the following reaction, $X_2~\rightarrow~X_2*$ with $R_2*$ as the enzyme.

$P_{1}~\{e12_1~\leftarrow~e9_2\}
~|[e9_2,~U]|~
P_{2}'$

 where $U=\{e2_1,~e3_1,~e4_1,~e7_1,~e8_1,~e9_1,~e10_1,~e14_1,~e3_2,~e4_2,~e5_2,~e7_2,~e8_2,~e10_2,~e12_2,~e14_2\}$\\
\indent and $P_{2}'~=~
Receptor_{2} ~ / ~ \{i1_{2}\} ~ \{e1_{2} ~ \leftarrow ~e5_{2},~e2_2~\leftarrow~e9_2\} ~
               |[e5_{2},~e9_{2},]|  ~ Cascade3_{2}~/~\{i2_{2}\}$\\
\indent\indent\indent\hspace*{0.9em}
$               \{e11_{2}~\leftarrow ~e13_{2}\}~|[e13_{2}]|  ~ GeneExpression_{2} $.

\paragraph{Receptor Function Cross-talk }

$P_{1}~|[E,~U]|~P_{2}$
has receptor function cross-talk if $E$ only contains labels from
 a receptor module and catalysis/inhibition labels from either a receptor module or a cascade module.
Example: provide an alternative route to activate receptor $R_2$ by the enzyme $X_1*$. Synchronise $e3_2$, the alternative route to active $R_2$, with $e8_1$, the enzymatic activity of $X_1*$.

$P_{1}~\{e8_1~\leftarrow~e3_2\}~|[e3_2,~U]|~P_{2}$

where $U=\{e2_1,~e3_1,~e4_1,~e7_1,~e9_1,~e10_1,~e12_1,~e14_1,~e2_2,~e4_2,~e7_2,~e8_2,~e9_2,~e10_2,~e12_2,~e14_2\}$.

\paragraph{Gene Expression Cross-talk }

$P_{1}~|[E,~U]|~P_{2}$
has gene expression cross-talk if $E$ only contains labels from
a gene expression module and catalysis/inhibition labels from a cascade module.  Example: 
connect the inhibiting activity of $Z_2*$  to the expression of $Gene_1$.
Synchronise $e10_2$, the inhibiting activity of $Z_2*$, with $e13_1$, the
expression of $Gene_1$.

$P_{1}~|[e13_1,~U]|~P_{2}~\{e10_2~\leftarrow~e13_1\}$

where $U=$

\indent\indent $\{e2_1,~e3_1,~e4_1,~e7_1,~e8_1,~e9_1,~e10_1,~e12_1,~e14_1,~e2_2,~e3_2,~e4_2,~e7_2,~e8_2,~e9_2,~e12_2,~e14_2\}$.

\paragraph{Intracellular Communication Cross-talk}

$P_{1}~|[E,~U]|~P_{2}$
has intracellular communication cross-talk if $E$ only contains a protein degradation label from
 a gene expression module and a ligand production  label from a receptor module. Example
connect the degradation  of $Protein_1$  to the production of $L_2$.
Synchronise $e14_1$, the degradation of  $Protein_1$, with $e4_2$, the
production of $L_2$.

$P_{1}~\{e14_1~\leftarrow~e4_2\}~|[e4_2,~U]|~P_{2}$ 

where $U=$ $\{e2_1,~e3_1,~e4_1,~e7_1,~e8_1,~e9_1,~e10_1,~e12_1,~e2_2,~e3_2,~e7_2,~e8_2,~e9_2,~e10_2,~e12_2,~e14_2\}$.

\subsection{Detecting Cross-talk}

We perform Continuous Stochastic Logic (CSL) model checking in PRISM to detect the presence of cross-talk.  
For each of the 5 cross-talk models we compute the probability of three properties
and note that cross-talk is detected by a change in probability 
compared with the independent model.
The three properties are as follows.
\\

\emph{Competitive Signal Flow ($P_1$):  }
 probability of  signal  flow through $P_1$  before $P_2$

{\small{
\
\hspace*{2em}
$\mathbf{P_{=?}}~[~F(Protein_1~=~1~\wedge~Protein_2~=~0)~]$}
}     
\\

\emph{ Time-dependent Signal Flow ($P_1$):}
 probability of  signal  flow through $P_1$ within 3 time units

{\small{
\
\hspace*{2em}
$\mathbf{P_{=?}}~[~F_{\leq 3}(Protein_1~=~1)~]$}
}
\\

\emph{Time-dependent Signal Flow ($P_2$): }
 probability of  signal  flow through $P_2$ within 3 time units

{\small{
\
\hspace*{2em}
$\mathbf{P_{=?}}~[~F_{\leq 3}(Protein_2~=~1)~]$}
}
\\          

\begin{center}
\begin{tabular}{|l|c|c|c|}
\hline
& Competitive & Time-dependent & Time-dependent \\
& Signal Flow ($P_1$) & Signal Flow ($P_1$) & Signal Flow ($P_2$)\\
\hline
\hline
Independent Pathways & $0.500$ & $0.184$ &  $0.18473$ \\ 
\hline
\hline
Signal Flow Cross-talk  & \multirow{2}{*}{$0.638$} & \multirow{2}{*}{$0.304$} & \multirow{2}{*}{$0.18473$}\\
(flow from $P_2$ to $P_1$)  &  &  & \\
\hline
Substrate Availability Cross-talk & \multirow{2}{*}{$0.500$} & \multirow{2}{*}{$0.141$} &   \multirow{2}{*}{$0.14125$} \\
($P_1$ and $P_2$ compete for a protein)  &  &  & \\
\hline
Receptor Function Cross-talk & \multirow{2}{*}{$0.487$} & \multirow{2}{*}{$0.184$}  & \multirow{2}{*}{$0.19257$} \\       
($P_1$ activates $P_2$'s receptor)  &  &  & \\
\hline
Gene Expression Cross-talk & \multirow{2}{*}{$0.363$} &  \multirow{2}{*}{$0.147$} & \multirow{2}{*}{$0.18473$} \\
($P_2$ inhibits $P_1$'s gene expression)  &  &  & \\
\hline
Intracellular Communication Cross-talk & \multirow{2}{*}{$0.500$} &   \multirow{2}{*}{$0.184$}  &   \multirow{2}{*}{$0.18477$} \\
($P_1$ expresses $P_2$'s ligand)  &  &  & \\
\hline
\end{tabular}  
\end{center}

The competitive signal flow property  detects the presence of 3 of the  5 types of cross-talk.  The probability of this property is significantly greater  with  signal flow cross-talk, due to the extra signal flow from $P_2$,
and significantly 
less with gene expression cross-talk, due to inhibition of gene expression by $P_2$.
There is a decrease in the probability of this property with receptor function cross-talk, however the decrease is small due to the likelihood that the cross-talk is initiated after $P_2$'s receptor becomes activated.
The competitive signal flow property does not detect 
the substrate availability cross-talk, as the cross-talk has an identical effect on each pathway. Intracellular communication cross-talk is also not detected, as the cross-talk occurs after $protein_1$ is expressed (the property does not refer to behaviour past this point).

To detect substrate availability and intracellular communication cross-talk we need to check the probability of independent signal flow through a pathway (hence, not in relation to another pathway).
This is accomplished by checking the probability of signal flow through a pathway within a time bound, in this case (arbitrarily) chosen to be 3 time units.
Substrate availability cross-talk  has the effect of equally decreasing the time-dependent signal flow through both $P_1$ and $P_2$ due to the competition for a limited protein.
Intracellular communication cross-talk has the effect of increasing only the signal flow through $P_2$, however the effect is marginal (5th decimal place).    It is interesting to note that in Section \ref{cellCommSection} we identified intracellular communication cross-talk as a source of contention amongst the community and it proves difficult to detect in our analysis.

\subsection{Characterising Cross-talk}
	  
We now define 5  CSL properties, each of which characterises a type of cross-talk and thus holds only in the respective model defined previously.
The properties are simple liveness or safety  properties and do not exploit the rate information in the model.  These properties could equally be written in Computational Tree Logic, replacing the probabilistic operator with the universal (A) and existential (E) operators as appropriate.

\paragraph{Signal Flow Cross-talk  (flow from $P_2$ to $P_1$)}

it is possible to activate $P_1$ without activating receptor $R_1$ 

{\small{              
$\mathbf{P_{> 0}}~[~F(RActive_1~=~0~\wedge~Protein_1~=~1)~]$}
}

\newpage

\paragraph{Substrate Availability Cross-talk
($P_1$ and $P_2$ compete for a protein) }

it is not possible to activate both $P_1$ and  $P_2$  (i.e. the pathways compete for a limited protein)

{\small{     
$\mathbf{P_{\leq 0}}~[~F(Protein_1~=~1~\wedge~Protein_2~=~1)~]$}
}

\paragraph{Receptor Function Cross-talk     
($P_1$ activates $P_2$'s receptor) }

it is possible to activate the receptor $R_2$ without using the ligand $L_2$

{\small{      
$\mathbf{P_{> 0}}~[~F(RActive_2~=~1~\wedge~L_2~=~1)~]$}
}

\paragraph{
Gene Expression Cross-talk 
($P_2$ inhibits $P_1$'s gene expression) }

it is not possible to express $protein_1$ if the signal has already passed through $P_2$

{\small{    
$\mathbf{P_{\leq 0}}~[~F(Protein_1~=~1)~\{Protein_1~=~0~\wedge~Protein_2~=~1\}~]$}
}      

\paragraph{
Intracellular Communication Cross-talk 
($P_1$ expresses $P_2$'s ligand)}

it is possible to use and replenish ligand $L_2$

{\small{
$\mathbf{P_{> 0}}~[~(L_2~=~1)~\wedge~(L_2~=~1)~U~(
~(L_2~=~0)~\wedge~(L_2~=~0)~U~(L_2~=~1)~)~]$}
}

\section{Case Study:  TGF-$\beta$/BMP, WNT and MAPK pathways}
\label{case-study}

We applied our approach to a prominent  biological case study of the cross-talk      between the TGF-$\beta$/BMP, WNT and MAPK pathways.
Details are taken from \cite{guo2009}  and  from discussions with a domain expert \cite{rainerCorresp}. 
We consider the behaviour of the independent pathways and the various types of cross-talk.  
We note that the effects of cross-talk are not discussed in \cite{guo2009}.

\begin{figure}[ht]
\centering                     
\includegraphics[trim=0in  0in 0in 0in, clip,  
angle=-90,width=6in]{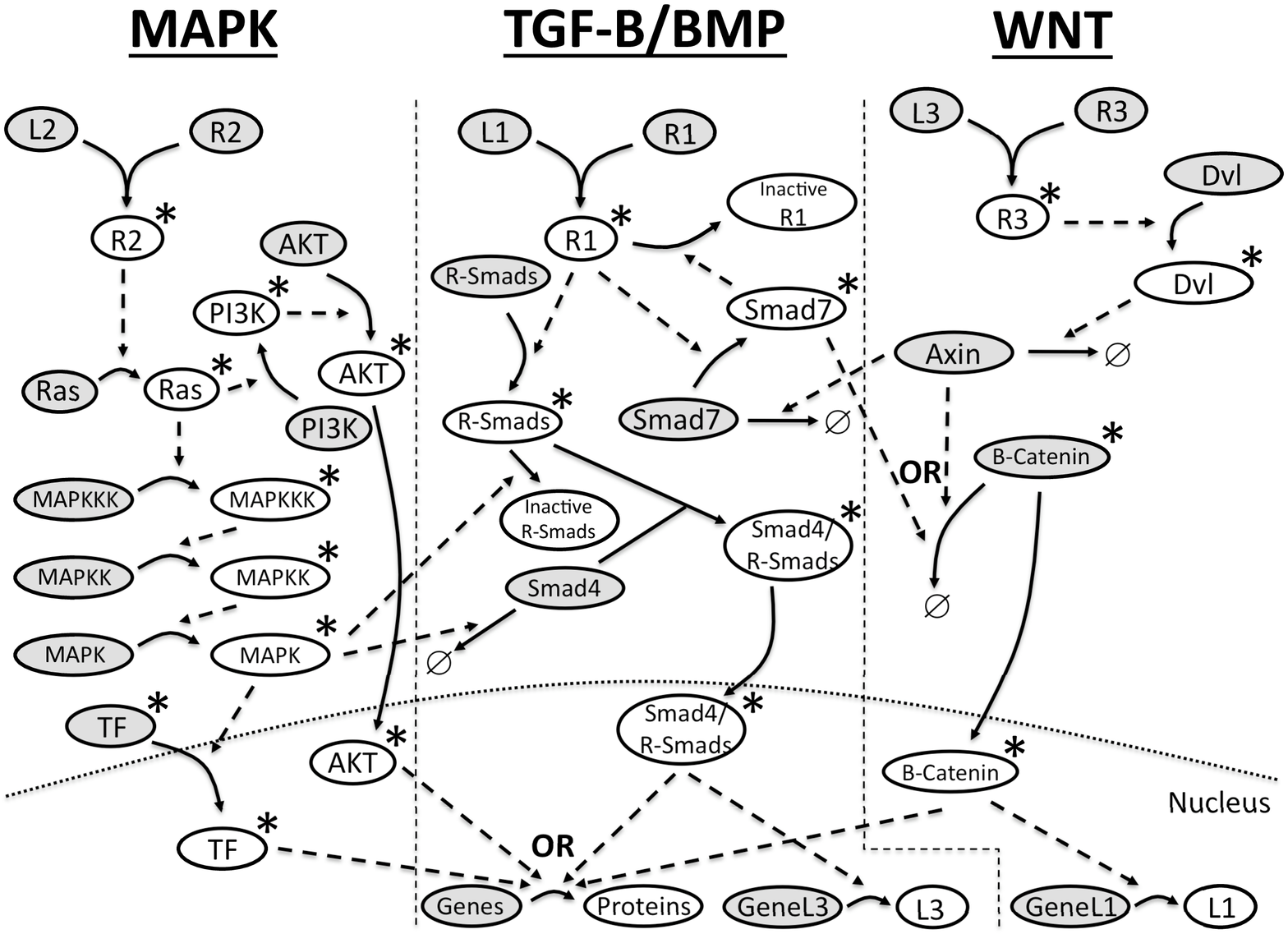}\\
\caption{Cross-talk between the TGF-$\beta$/BMP, WNT and MAPK pathways.
   Species that are  present in the initial state are denoted by shaded ellipses.
\label{TGF}} 
\end{figure}

Our model of the  pathways and their cross-talk is shown in  Figure \ref{TGF}.
To apply our module approach  we need to expand our set of modules to:  receptor, protein activation, 2-stage cascade, 3-stage cascade, translocation, protein binding and gene expression.
This is a natural extension of our approach, and the extra modules act in a similar manner to the modules that have been discussed.  
Using the formal descriptions of cross-talk  
(Section \ref{examples}), we have identified three types of cross-talk in this model.

We measure the output of the TGF-$\beta$/BMP pathway  by the activity of the expression of Genes to Proteins.  
 We use the following properties to compare the effects of  cross-talk: $\psi_1$, the eventual  expression of Genes,  and $\psi_2$, the time-dependent  expression of Genes (within 5 time units). 

{\small{
\
\hspace*{2em}
$\psi_1~=~\mathbf{P_{=?}}~[~F(Proteins~=~1)~]$, \indent $\psi_2~=~\mathbf{P_{=? }}~[~F_{\leq 5}(Proteins~=~1)~]$}
}     

\paragraph{Independent Pathways  }
With independence,
the activation of the TGF-$\beta$/BMP pathway leads to gene expression within 5 time units, $\psi_2$,  with probability $0.47$ and eventual gene expression, $\psi_1$, with probability $<$ $1$ due to the inactivation of the receptor.

\paragraph{TGF-$\beta$/BMP and MAPK Cross-talk}  
There are two types of  cross-talk between the TGF-$\beta$/BMP and MAPK pathways.  
\emph{Signal flow:}   $MAPK*$ proteins slow signal flow through the TGF-$\beta$/BMP pathway by deactivating the $R$-$Smads$ and  degrading  $Smad4$.
\emph{Gene expression:}  the $TF*$ and $AKT*$ proteins upregulate  gene expression  in the TGF-$\beta$/BMP pathway. 
Note that the appearance of the $AKT$ and $PI3K$ proteins in the MAPK pathway indicates an implicit cross-talk with the  $AKT$ and $PI3K$ pathways respectively.
The inclusion of cross-talk with the MAPK pathway can both provide alternative gene expression routes and block the TGF-$\beta$/BMP route, overall causing  the probability of gene expression within 5 time units,  
 $\psi_2$, to marginally decrease   to  $0.73$.  
The probability of eventual gene expression, $\psi_1$, is $1$ due to the  consistent routes  through the MAPK pathway.

\paragraph{TGF-$\beta$/BMP and WNT Cross-talk}

There are three types of  cross-talk between the TGF-$\beta$/BMP and WNT pathways.  
\emph{Signal flow:}   the $Smad7*$ protein degrades $\beta$-$Catenin$ and the $axin$ protein degrades $Smad7$\footnote{this cross-talk is discussed further in Section \ref{discussion}}.
\emph{Gene expression:} the $\beta$-$Catenin$ protein upregulates gene expression in the TGF-$\beta$/BMP pathway.
\emph{Intracellular communication:}    the WNT pathway can cause the production of a ligand for the  TGF-$\beta$/BMP pathway, and vice-versa.
The inclusion of cross-talk with the WNT pathway can both provide an alternative route to gene expression and inhibit $Smad7$ which can inactivate the receptor for the  TGF-$\beta$/BMP pathway.  Overall this causes the probability of gene expression within 5 time units,  
 $\psi_2$, to marginally increase to  $0.76$.  
The probability of eventual gene expression, $\psi_1$, is still $<$ $1$ due to the degradation of the $\beta$-$Catenin$ protein.

\paragraph{TGF-$\beta$/BMP, WNT and MAPK Cross-talk}   
The cross-talk between all three pathways is the union of the two cross-talk scenarios  above.  
The effect of both WNT and MAPK cross-talk to the TGF-$\beta$/BMP pathway is additive.  The probability of $\psi_2$ has risen to $0.88$, compared with
 the single cross-talks of WNT  and MAPK with probability  $0.76$ and $0.73$  respectively.  
The inclusion of the MAPK cross-talk provides consistent routes to gene expression and hence the probability of  $\psi_1$ is $1$.

\section{Discussion}    
\label{discussion}

\paragraph{Categorisation}   
The categorisation of signal flow   cross-talk between the WNT and TGF-$\beta$/BMP pathways in the previous section may be open to discussion.  
In Figure \ref{TGF}  $Axin$ degrades $Smad7$.  Rather than being part of the  signal flow from the TGF-$\beta$/BMP receptor to gene expression,  $Smad7$ actually deactivates the TGF-$\beta$/BMP receptor.  One could consider this cross-talk as receptor function cross-talk with an intermediate ($Smad7$).  Our approach categorises this as signal flow because  the set of labels involved is an catalysis label from a protein module and a degradation label from a protein activation module.

\paragraph{Quantitative Detail}
We have demonstrated our approach in this paper on models with a low level of  quantitative detail.
As such, the probability values resulting from CSL model checking can only be used to compare between models. 
However, with more quantitative detail, further interpretation of our analysis results would be possible.                    
For example, the properties concerning the probability of time-dependent   gene expression between cross-talk models would become a meaningful assessment of the strength of the cross-talk.

\paragraph{Feature Interaction}
There may be an interesting analogy with feature interactions in telecommunications and software systems.  Features, or services, in these systems are additional functionality (additional to the core). They are often added incrementally, by various developers, at various times (e.g. due to deregulation).  A possible consequence is interactions between the new features themselves, or with the core system,  causing some features or the core to behave in new, sometimes undesirable ways.  An open question is whether techniques developed to model and detect features and interactions may be applicable to pathway cross-talk. Moreover, a common problem is lack of universal definition of pathway/feature; it would be interesting to investigate if concepts such as the feature construct of \cite{plry2} would be useful in the pathway paradigm.  We note that many approaches to interaction detection are based on temporal logic descriptions of behaviour.

\section{Related Work} 
\label{related}    
                  
The literature  contains a limited number of applications of computational techniques  to the study of signalling pathway cross-talk.

There are several examples whereby a proposed method of cross-talk between two pathways is expressed and analysed as a computational model.  For example \cite{hatakeyama} and \cite{sreenath} use Ordinary Differential Equations (ODEs) to model the cross-talk between the MAPK pathway with the AKT and PKC pathways respectively.                              
The work in \cite{Bauer-net-2009} is similar, however  the modelling technique used to model the cross-talk between the growth factor pathway and the Integrin pathway is   stochastic networks.
Furthermore, a more formal notation of Petri nets has been used in \cite{heinerPetriCrosstalk} to model the cross-talk between the pathways involved in Apoptosis decision-making.             
     
The computational analysis of cross-talk models have produced some interesting results in the literature.  
For example, \cite{citeulike:1097014} has produced a model containing substrate availability cross-talk between the hyperosmolar and the pheromone MAPK pathways.  The question the authors try to answer is how these pathways maintain signal specificity given that they share common proteins.  The authors analyse two models which can account for the signal specificity, one   which contains mutual inhibition between pathways to limit signal bleed-through and one  which contains scaffold proteins.   
In \cite{sreenath} the computational analysis shows that cross-talk has an effect on whether the model exhibits  bistability.  With pathway cross-talk, the behaviour of key proteins  switches from transient to sustained activation upon varying the duration of the signal.

We have found only one paper on the application of formal models to cross-talk, \cite{citeulike:1309044}, which contains a model of the multiple modes of cross-talk between the EGFR and LIN-12/Notch signalling pathways.  A discrete, dynamic, state-based  model  is developed in using the language of Reactive Modules.
Model checking is used to check the validity of the model and to generate  ``new biological insights into the regulatory network governing the cell fate''.  
However, this work  differs from ours because it concerns
intercellular cross-talk within a    multi-cellular model.  
The application of the term cross-talk to intercellular communication is often considered a misnomer amongst the community, and hence we have focussed our analysis to intracellular cross-talk in a single-cell model.

\section{Conclusion and Future Work}
\label{conclusions}

In this paper we have explored  formal modelling and analysis techniques for signalling pathway cross-talk.  We note that formal methods, whilst applied frequently in the literature to signalling pathways, have been largely ignored in the study of cross-talk.  The aim of this paper has been to show that formal methods are both  well-suited to and a natural choice for this area.

Our first  contribution is a novel categorisation of cross-talk, drawing from examples in the literature. 

The second contribution is the definition of a pathway and pathway composition, based on a set of generic pathway modules with  internal and external reactions, and renaming and synchronisation operations. We can compose pathway independently, or with cross-talk. 
Furthermore, we also find that temporal logic descriptions of behaviours are suitable to detect and characterise cross-talk. The approach is illustrated with example pathways. 

The third and final contribution is the application  to cross-talk between the TGF-$\beta$/BMP pathway and two other pathways. 

Several future directions have been identified.  As suggested earlier in the paper, we wish to apply our approach to models with a higher level of quantitative detail to make predictions and generate insights about the biological effects of cross-talk.  Furthermore, we wish to assess how the effect of cross-talks differs to that of standard pathway motifs.  Hence, is there a reasonable alteration of the pathway model (e.g. addition of a feedback loop) that gives the same behaviour as a potential cross-talk?
We also wish to assess how the effectiveness of pathway intervention techniques  (e.g. drugs, gene knockouts) changes with the addition of cross-talk.  
Finally, a larger question is how the temporal ordering of signals affects the  detectability and behaviour of cross-talk (i.e. do the pathways hold a    ``biochemical history'' of signalling events?).

The PRISM models and CSL properties used in this paper can be found at:

\url{www.dcs.gla.ac.uk/~radonald/fbtc2010/}.

\subsubsection*{Acknowledgements}

This research is supported by the project \emph{The Molecular Nose}, funded by the Engineering and Physical Sciences Research Council (EPSRC).
We wish to thank Rainer Breitling for the collaboration on the development of the case study model.

\bibliography{biblio}

\begin{thebibliography}{10}

\bibitem{brightman2000}
Brightman, F., Fell, D.:
\newblock {Differential feedback regulation of the MAPK cascade underlies the
  quantitative differences in EGF and NGF signalling in PC12 cells}.
\newblock FEBS Lett 482 (2000)  169--174

\bibitem{citeulike:479006}
Taniguchi, C.M., Emanuelli, B., Kahn, R.C.:
\newblock Critical nodes in signalling pathways: insights into insulin action.
\newblock Nat Rev Mol Cell Biol \textbf{7}(2) (February 2006)  85--96

\bibitem{catt1967}
Catt, I.:
\newblock {Crosstalk (Noise) in Digital Systems}.
\newblock Electronic Computers, IEEE Transactions on \textbf{EC-16}(6) (1967)
  743--763

\bibitem{citeulike:1198248}
Schwartz, M.A., Ginsberg, M.H.:
\newblock Networks and crosstalk: integrin signalling spreads.
\newblock Nat Cell Biol \textbf{4}(4) (April 2002)  E65--E68

\bibitem{guo2009}
Guo, X., Wang, X.F.:
\newblock {Signaling cross-talk between TGF-$\beta$/BMP and other pathways}.
\newblock Cell Research \textbf{19} (2009)  71--88

\bibitem{PA-07-058}
{NIH}:
\newblock {``Insulin Signaling And Receptor Cross-Talk'' Program Announcement
  PA-07-058. November 20, 2006.}

\bibitem{citeulike:1097014}
McClean, M.N.N., Mody, A., Broach, J.R.R., Ramanathan, S.:
\newblock {Cross-talk and decision making in MAP kinase pathways.}
\newblock Nat Genet (January 2007)

\bibitem{estrogenReceptor}
Katzenellenbogen, B.:
\newblock Estrogen receptors: bioactivities and interactions with cell
  signaling pathways.
\newblock Biol. Reprod. \textbf{54} (1996)  287--293

\bibitem{bosscher2006}
Bosscher, K.D., Berghe, W.V., Haegeman, G.:
\newblock Cross-talk between nuclear receptors and nuclear factor $\kappa$b.
\newblock Oncogene \textbf{25} (2006)  6868--6886

\bibitem{orton2005}
Orton, R.J., Sturm, O.E., Vyshemirsky, V., Calder, M., Gilbert, D.R., Kolch,
  W.:
\newblock {Computational modelling of the receptor tyrosine kinase activated
  MAPK pathway}.
\newblock Biochemical Journal \textbf{392}(2) (2005)  249--261

\bibitem{www:prism}
{PRISM Website}:
\newblock {PRISM - Probabilistic Symbolic Model Checker}.
\newblock {\url{http://www.prismmodelchecker.org}}

\bibitem{2005cbio}
Calder, M., Vyshemirsky, V., Orton, R., Gilbert, D.:
\newblock {Analysis of Signalling Pathways using Continuous Time Markov
  Chains}.
\newblock Trans. on Computat. Syst. Biol. VI \textbf{4220} (2006)  44--67

\bibitem{ctmcLevels}
Ciochetta, F., Degasperi, A., Hillston, J., Calder., M.:
\newblock {Some investigations concerning the CTMC and the ODE model derived
  from Bio-PEPA}.
\newblock Proceedings of FBTC 2008, ENTCS \textbf{17350} (2008)

\bibitem{rainerCorresp}
Breitling, R.:
\newblock Private correspondence

\bibitem{plry2}
Plath, M., Ryan, M.:
\newblock The feature construct for {SMV}: {S}emantics.
\newblock In Calder, M., Magill, E., eds.: Feature Interactions in
  Telecommunications and Software Systems VI, IOS Press (Amsterdam) (2000)
  129--144

\bibitem{hatakeyama}
Hatakeyama, M., Kimura, S., Naka, T., Kawasaki, T., Yumoto, N., Ichikawa, M.,
  Kim, J., Saito, K., Saeki, M., Shirouzu, M., Yokoyama, S., Konagaya, A.:
\newblock {A computational model on the modulation of mitogen-activated protein
  kinase (MAPK) and Akt pathways in heregulin-induced ErbB signalling.}
\newblock Biochemical Journal \textbf{373 Pt. 2} (2003)  451--463

\bibitem{sreenath}
Sreenath, S.N., Soebiyanto, R., Mesarovic, M.D., Wolkenhauer, O.:
\newblock {Coordination of crosstalk between MAPK-PKC pathways: an exploratory
  study}.
\newblock Systems Biology, IET \textbf{1} (2007)  33--40

\bibitem{Bauer-net-2009}
Bauer, A., Jackson, T., Jiang, Y., Rohlf, T.:
\newblock Stochastic network model of receptor cross-talk explains pro- and
  anti-angiogenic control.
\newblock under review (2009)

\bibitem{heinerPetriCrosstalk}
Heiner, M., Koch, I., Will, J.:
\newblock {Model validation of biological pathways using Petri
  nets--demonstrated for apoptosis.}
\newblock Journal BioSystems \textbf{75} (2004)  15--28

\bibitem{citeulike:1309044}
Fisher, J., Piterman, N., Hajnal, A., Henzinger, T.A.:
\newblock {Predictive Modeling of Signaling Crosstalk during C. elegans Vulval
  Development}.
\newblock PLoS Computational Biology \textbf{3}(5) (May 2007)  e92+

\end{thebibliography}

\end{document}